\documentclass{article}

\PassOptionsToPackage{compress, numbers}{natbib}     

\usepackage[preprint]{neurips_2023}

\usepackage[utf8]{inputenc} 
\usepackage[T1]{fontenc}    
\usepackage{hyperref}       
\usepackage{url}            
\usepackage{booktabs}       
\usepackage{amsfonts}       
\usepackage{nicefrac}       
\usepackage{microtype}      
\usepackage{xcolor}
\usepackage{amsmath} 
\usepackage{graphicx}
\usepackage{wrapfig}

\title{Microscopy image reconstruction with physics-informed denoising diffusion probabilistic model}

\author{%
  Rui Li\textsuperscript{1,}\thanks{Equal contribution}, Gabriel della Maggiora\textsuperscript{1,*},\\
  \And
  Vardan Andriasyan\textsuperscript{2,3}, Anthony Petkidis\textsuperscript{2,3}, Artsemi Yushkevich\textsuperscript{4}, Mikhail Kudryashev\textsuperscript{4,5},
  \And
  Artur Yakimovich\textsuperscript{1,3}\\
 \textsuperscript{1}Center for Advanced Systems Understanding (CASUS),\\ Helmholtz-Zentrum Dresden-Rossendorf e. V. (HZDR), Görlitz, Germany\\
\textsuperscript{2}Department of Molecular Life Sciences, University of Zurich, Zurich, Switzerland\\
\textsuperscript{3}Artificial Intelligence for Life Sciences CIC, Dorset, United Kingdom\\
\textsuperscript{4}Max Delbrück Center for Molecular Medicine in the Helmholtz Association, Berlin, Germany\\
\textsuperscript{5}Institute of Medical Physics and Biophysics, Charite-Universitätsmedizin, Berlin, Germany\\
  \texttt{correspondence: a.yakimovich@hzdr.de} \\
}

\begin{document}

\maketitle

\begin{abstract}
Light microscopy is a widespread and inexpensive imaging technique facilitating biomedical discovery and diagnostics. However, light diffraction barrier and imperfections in optics limit the level of detail of the acquired images. The details lost can be reconstructed among others by deep learning models. Yet, deep learning models are prone to introduce artefacts and hallucinations into the reconstruction. Recent state-of-the-art image synthesis models like the denoising diffusion probabilistic models (DDPMs) are no exception to this. We propose to address this by incorporating the physical problem of microscopy image formation into the model's loss function. To overcome the lack of microscopy data, we train this model with synthetic data. We simulate the effects of the microscope optics through the theoretical point spread function and varying the noise levels to obtain synthetic data. Furthermore, we incorporate the physical model of a light microscope into the reverse process of a conditioned DDPM proposing a physics-informed DDPM (PI-DDPM). We show consistent improvement and artefact reductions when compared to model-based methods, deep-learning regression methods and regular conditioned DDPMs.

\end{abstract}

\section{Introduction}
Since its discovery, light microscopy (LM) remains an important and accessible way to explore the hidden biomedical world. The cost of simple LM equipment keeps dropping making them available to classrooms for education \cite{knapper2022fast,salido2022review,maia2017100,aidukas2019low} and medical laboratories for applications like cytometry \cite{knapper2022fast}. Furthermore, the contribution of advanced LM techniques like fluorescence microscopy \cite{conchello2005fluorescence}, confocal microscopy \cite{minsky1988memoir} or superresolution microscopy \cite{schermelleh2019super} to a plethora of biomedical discoveries of the past century \cite{nechyporuk2022principles} are hard to overstate.

Among the notable techniques widely used in laboratories and classrooms of the world, one could name upright and inverted brightfield microscopy, widefield fluorescence microscopy (epifluorescence) and confocal microscopy  \cite{nechyporuk2022principles,minsky1988memoir,conchello2005fluorescence}. In brightfield (transmission light) microscopy, the image is formed through direct interaction of the white illumination light with the specimen. In widefield microscopy image formation is a result of the excitation of the molecular fluorophores, which emit photons contributing to the image. Finally, confocal microscopy builds upon widefield fluorescence microscopy by removing the emission light coming from outside of the immediate focal plane, thereby reducing blur in the image. This is achieved by introducing a pinhole in the light path of the microscope \cite{nechyporuk2022principles}.

\section{Related work}
\subsection{Light Microscopy Point Spread Function Model}

\begin{wrapfigure}{r}{0.5\textwidth}
  \begin{center}
    \includegraphics[width=0.48\textwidth]{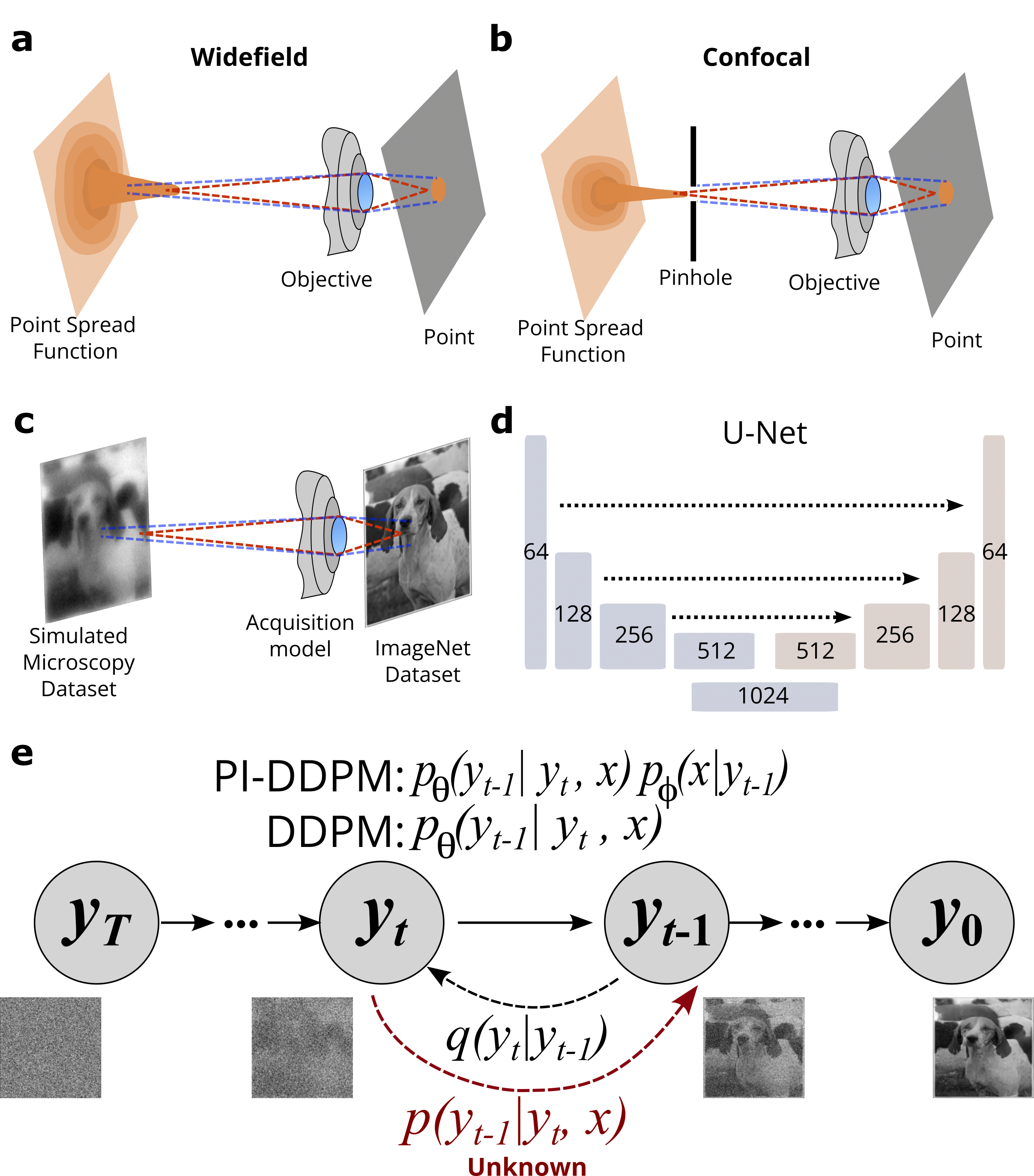}
  \end{center}
\caption{\begin{small}\textbf{Proposed model with physics-informed probabilistic denoising diffusion for Microscopy image reconstruction}. (a,b) simplified schematic depiction of widefield and confocal microscopy. (c) schematic depiction of synthetic dataset generation using our acquisition model. (d) schematic depiction of a U-Net architecture. (e) illustration of a denoising diffusion probabilistic model and the physics-informed version we propose.\end{small}}
\end{wrapfigure}

Yet, optical systems remain fundamentally limited owing to the principles of their design. With few exceptions, all LM systems obtain the image by collecting the light interacting with the specimen using a system of optical lenses. Passing through these components light becomes scattered and distorted causing imperfections and blur in the obtained image. The blur and imperfections of a point source (e.g. single molecule fluorescence) can be expressed mathematically as a point spread function (PSF) \cite{shaw1991point}. PSF describes the spread of light that occurs from scattering and diffraction as it passes through the optical components of the microscope. The advent of digital microscopy and image processing allowed us to attempt alleviating these limitations algorithmically in a process referred to as deconvolution \cite{richardson1972bayesian, nasse2010realistic, lucy1974iterative}. While these algorithms are capable of significant improvement in image quality, they are prone to introduce artefacts due to their simplicity. On the other side of the spectrum, recent advances in deep learning (DL) with trainable models like convolutional neural networks (CNNs) allow for data-driven image restoration \cite{weigert2018content, buchholz2019content, noh2015learning}. However, conventional trainable solutions require a large scale of training data, massive learning models and longer training time.

To address these shortcomings, we introduce a physics-based diffusion model with a physics-informed term incorporated into the loss function. We train our model on ImageNet \cite{deng2009imagenet} simulated to look like micrographs. We show that this approach not only provides a simpler and more principled model but also produces more natural-looking results.

The objective of image reconstruction in microscopy is to obtain the high-frequency details that are lost due to the diffraction limit and optical imperfections of a microscope. Mathematical methods used for image reconstruction in microscopy can be categorised as deconvolution methods \cite{sarder_deconvolution_2006, mcnally_three-dimensional_1999, shaw1991point}, regularisation methods \cite{lustig_sparse_2007, dutta_joint_2012,  tang_superpenetration_2012} and bayesian methods \cite{yoo_bayesian_2018}. Traditional methods, however, do not capture the complexity of the images leading to reconstructions that show little improvement in resolution. Additionally, these methods can be susceptible to noise and artefacts present in acquiring the images leading to degraded performance in the reconstructed image.

To address these issues, researchers have turned to DL models. These models have shown promising results in different tasks, particularly image reconstruction. For example, Xu and co-authors \cite{xu2014deep} proposed to use CNNs to capture the characteristics of degradation, rather than modelling outliers perfectly. Specifically, the authors transformed a simple pseudo-inverse kernel for deconvolution into a CNN. Later, Ronneberger and colleagues incorporated the deconvolution layers in their U-Net architecture \cite{ronneberger2015u}. More recently, the image restoration task has been attempted with generative adversarial networks \cite{saqlain2022dfgan, qiao2021evaluation}.

Another recent trend is a class of likelihood-based generative models called Denoising Probabilistic Diffusion Models (DDPMs) \cite{ho_denoising_2020}. They have shown promise in several tasks such as superresolution \cite{li_srdiff_2021}, image colourisation, inpainting, uncropping, and JPEG restoration \cite{saharia_palette_2022}. Furthermore, diffusion models have desirable properties such as distribution coverage, stationary optimisation function, scalability, and training stability and have shown better quality sample generation compared to generative adversarial networks\cite{dhariwal_diffusion_2021}.

However, most of these approaches largely ignore the great body of knowledge gathered on microscopy systems throughout the centuries by optical physics. Furthermore, one common issue with DL methods, especially generative models, is the tendency to generate unrealistic structure that is not present in the real image, which is problematic in fields in which an accurate reconstruction is important such as medical diagnostic images and microscopy \cite{bhadra_hallucinations_2021, uzunova_systematic_2022}. To circumvent this, in other domains where the physics of the process is well understood, researchers have recently proposed an approach called physics-informed neural networks \cite{raissi2017physics}. In this approach, prior knowledge of laws of physics may be employed as a regularisation for DL models. The method we propose here incorporates a physics-informed term into the loss function of DDPMs, particularly we propose to incorporate the physical problem into the DDPM loss function using the technique shown in \cite{dhariwal_diffusion_2021, ho_video_2022}.

\section{Methods}

In LM, the diffraction pattern generated in an ideal optical system is the impulse response referred to as the point spread function (PSF) \cite{shaw1991point}. PSF of LM varies depending on the specifics of the technique employed, e.g. widefield and confocal LM (Fig. 1a,b). The diffraction pattern generated in an ideal optical system is the impulse response referred to as the point spread function (PSF). In fluorescence microscopy, usually, the illumination (excitation) and detection (emission) wavelength are not the same, so the most suitable model of the PSF \cite{pawley2006handbook} can be expressed as:

\[h(x, y, z) = |u_{\lambda_\text{ex}}(x,y,z)|^2|u_{\lambda_\text{em}}(x,y,z)|^2,\]

where \(u_\lambda\) corresponds to the PSF function for a respective emission or excitation wavelength \(\lambda\). This model is known as the Airy diffraction pattern \cite{BornWolf:1999:Book}. To add the effect of the pinhole used in confocal microscopy, we can convolve a disk function with the pupil of the emission sample. The disk function is usually modelled as: 
\[T(x) = 1_{R^2\le X^2+Y^2},\]
where R is the radius of the pinhole. Using this notion, we can rewrite the PSF as:

\[h(x,y,z) = |T(x,y) * u_{\lambda_\text{em}}(x,y,z)|^2 |u_{\lambda_\text{ex}}(x,y,z)|^2.\]

To model \(u\) we follow the Arnison-Sheppard approach \cite{arnison20023d}. In this approach, we model the optical transfer function (OTF). In the Arnison-Sheppard model, the OTF is expressed as the autocorrelation of the pupil function. Mathematically, the OTF \(C(\vec{K})\) is expressed in the k-space:

\[C(\vec{K}) = \iiint Q(\vec{m}+\frac{1}{2}\vec{K})\cdot Q^*(\vec{m}-\frac{1}{2}\vec{K})d\vec{m},\]
where Q corresponds to the complex vectorial pupil function \cite{sheppard_vectorial_1997}.

The complex vectorial pupil function is a complex-valued function \(Q(\vec{m})\) of the position vector \(\vec{m} = (k_x,k_y,k_z)\) within the aperture of an optical system. This function can be expressed as:
\[Q(\vec{m}) = A(\vec{m})e^{i\phi(\vec{m})},\]

where \(A\) is the amplitude transmission function with respect to the numerical aperture of the microscope and \(\phi\) is the phase shift produced by aberrations and microscope imperfections. Finally, to obtain \(u\) the inverse Fourier transform is applied to the OTF. 

\subsection{Light Microscopy Image Acquisition Model}
LM is an imaging technique in which an analogue-digital converter detects the electrical impulses generated by the light.  As a result, image statistics can be  well represented by a Poisson process. If several acquisitions are made and averaged, then according to the central limit theorem, the statistics of the image can be modelled by the Gaussian process. The mathematical model for optical systems assumes that the model is linear and time-invariant. Therefore the image acquisition model is described by the equation \(I = \phi(h*x + b)\), where the image \(I\), is the result of the convolution between object \(x\) and system PSF \(h\) with the background signal noise \(b\) added. The Poisson noise \(\phi\) is applied afterwards over the true signal, given by the previous equation.  

\subsection{Simulated Dataset Generation}
\begin{wrapfigure}[35]{r}{0.5\textwidth}
  \begin{center}
    \includegraphics[width=0.48\textwidth]{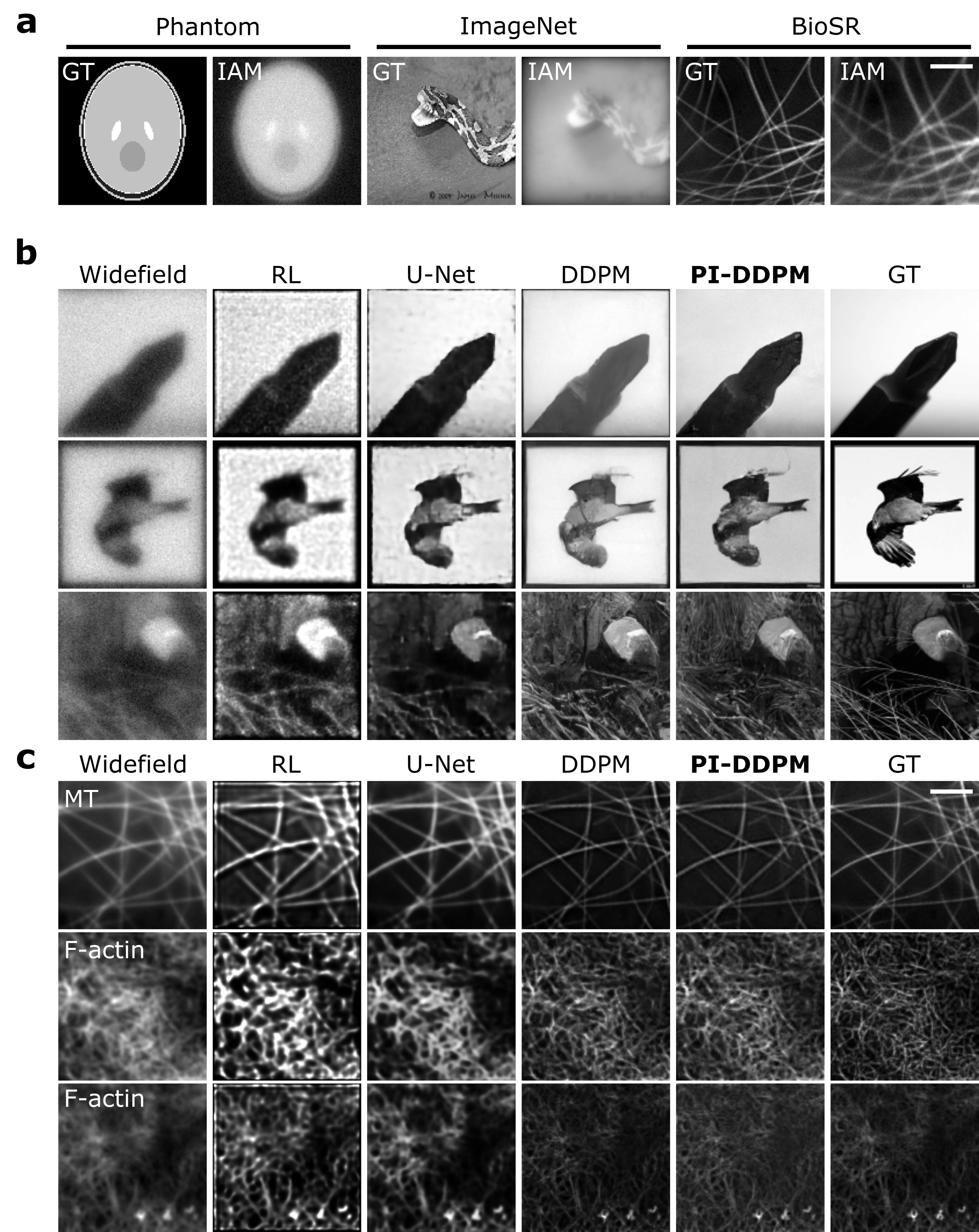}
  \end{center}
\caption{\begin{small}\textbf{Comparison of model performance on ImageNet and BioSR-derrived datasets}. (a) Examples of processing performed using Image Acquisition Model (IAM) on (left-to-right) Shepp–Logan phantom, ImageNet and BioSR images. (b, c) Examples of input (Widefield) and reconstructed images using Richardson-Lucy (RL), U-Net, Denoising Diffusion Probabilistic Models (DDPM) and physics-informed DDPM (PI-DDPM) in ImageNet and BioSR-derived images respectively. GT stands for ground truth. MT stands for microtubules. Scale bar in micrographs is 1.5 \(\mu\)m.\end{small}}
\end{wrapfigure}
We used the previously described diffraction model to simulate the effects of different microscopes over two sets of images: photographs obtained from ImageNet \cite{deng2009imagenet}, and structured illumination microscopy images obtained from the BioSR dataset \cite{qiao2021evaluation}. To mimic microscopy images using these datasets, each image of the dataset is convolved by a PSF and then had Poisoin noise applied to it (Fig. 1c). PSF for each image was randomly generated with parameters corresponding to physically plausible parameters for microscopy systems. Specifically,  we sampled numerical aperture between 0.4 and 1.0, the excitation wavelength between 320 \(\mu\)m and 400 \(\mu\)m, and the emission wavelength between 450 and 550 \(\mu\)m. The Pinhole size was sampled between 0.1 \(\mu\)m and 1000 \(\mu\)m. The focal plane was considered to be the centre of the volume object and the refractive index chosen was 1.33 which corresponds to the air refractive index. Using these parameters we generated 30000 different PSFs and randomly convolved between the object and the PSF to generate the  training dataset.

\subsection{Conditioned Denoising Probabilistic Diffusion models}
Denoising Probabilistic Diffusion Models (DDPMs) are latent variable models that use a sequence of latent variables to model the data \cite{ho_denoising_2020} (see Supplementary Materials). In the conditioned case, samples are drawn from an unknown distribution \(p(y|x)\). This is referred to as distribution because conditioned image synthesis is by nature ill-posed, specifically, there are many possible solutions \(y\) for any given input \(x\). In the conditioned process we want to learn an approximation of this distribution. To achieve that, it is possible to condition DDPMs in two ways.

The first way is to redefine the Markov chain. Given an image \(y\) and some corruption process \(p(x|y)\), we want to learn \(p(y|x)\). To achieve this, the diffusion Markov chain states are concatenated by the respective conditioning image \(x\) \cite{saharia_palette_2022}. Specifically, the distributions of the states of the Markov chain are generated by the diffusion process \(q(y_t|y_{t-1})\)
and concatenated to the conditioning image \(x\). Where \(y_i = \sqrt{\Bar{\alpha_t}}y_0 + \epsilon\sqrt{1-\Bar{\alpha_t}}, \quad \epsilon\sim \mathcal{N}(\epsilon;0,I)\).  To learn the reverse process, a reverse Markov chain is established, where \(p(y_T) = \mathcal{N}(y_T; 0,I)\):

\[p_\theta(y_{t-1}|y_t, x) = \mathcal{N}(y_{t-1}| \mu_\theta(x,y_t, \bar{\alpha_t}), \sigma^2I).\]

Using the same formulation as in regular DDPMs\cite{saharia_palette_2022}, we train the model to predict \(\epsilon\) at each time step:
\[\min_\theta L_\text{simple}:=\mathbb{E}_{t,(x,y_0),\epsilon}||\epsilon - \epsilon_\theta(y_t,x,\Bar{\alpha_t})||_2^2.\]

Finally, to obtain \(y_0\), the same iterative denoising as in the regular DDPMs is applied.

\[y_{t-1} = \frac{1}{\sqrt{\alpha_t}}\Bigg(y_t - \frac{\beta_t}{\sqrt{1-\bar{\alpha_t}}}\epsilon_\theta(x, y_t, \bar{\alpha_t})\Bigg) + \sqrt{\beta_t}\epsilon_t.\]

\subsection{Model-guided Denoising Probabilistic Diffusion models}
The second way to obtain conditioned samples from a diffusion model is to condition an unconditioned reverse process \cite{dhariwal_diffusion_2021}. Given an unconditional reverse process \(p_\theta(y_{t-1}|y_t)\), to condition on label \(x\) we can factorise \[p_{\theta,\phi} (y_{t-1}|y_t, x) = Zp_\theta(y_{t-1}|y_t)p_\phi(x|y_{t-1}),\]
where \(Z\) is a normalizing constant. This expression can be approximated as a perturbed Gaussian distribution. Since our unconditioned reverse process is a Gaussian, we have:

\[p_\theta(y_{t-1}|y_t) = \mathcal{N}(\mu, \sigma)\]
\[\log p_\theta(y_{t-1}|y_t) = -\frac{1}{2} (y_{t-1} - \mu)^T\Sigma^{-1}(y_{t-1} - \mu)  + C.\]

Since, at infinity, the distribution of the reverse process tends to a delta distribution, then it is reasonable to approximate \(p(x|y_t)\) by its Taylor expansion around the mean. 

\[\log p(x|y_t) \approx \log p(x|y_t)|_{y_t = \mu} + (y_t - \mu)\nabla_{y_t} \log p(x|y_t)|_{y_t=\mu} = (y_t - \mu)g + C_1\]

\centerline{where \(g  = \nabla_{y_t} \log p(x|y_t)\)}

Finally, by replacing and rearranging, we get:
\[p(y_{t-1}|y_t)p(x|y_{t-1}) \approx \mathcal{N}(\mu + \Sigma g, \Sigma).\]

Thus, the reverse conditioned process approximates the unconditioned Gaussian transition with its mean shifted by \(\Sigma g\).

\subsection{Physics-informed Denoising Probabilistic Diffusion Models}

Physics-informed methods arise from the need to incorporate physical knowledge into the construction or training of the model. In inverse problems, the objective is to determine \(p(y|x)\) where \(y\) is the true object and \(x\) is some observation of the object that follows a possibly stochastic process \(p(x|y)\).  In image reconstruction problems such as MRI undersampled image reconstruction, deblurring or microscopy image reconstruction, the physical model is generally of the form \(x = \phi(K*y)\) where \(\phi\) is a function that applies noise according to a certain distribution. Particularly, model-based methods try to solve a problem of the form \[\min_y ||K*y - x||^2 + \lambda R(y).\] Where \(\lambda\) is the weight of the regulariser \(R(y)\). To recover the true object from the observed variables. This formulation can be incorporated readily into the model guided DDPM, to obtain improvements over the conditioned case. However, one cannot access \(K\) during inference in several cases. To remediate this, we suggest learning the shifted mean of a conditioned DDPM by the physical model. In this sense, we incorporate the gradient of the solution of the physical problem as the shift of the mean of the unconditioned case.

Given \(K\),\(x\), \(\tilde{x}\) and \(y_t\), where \(x\) is the noiseless version of \(\tilde{x}\). Using the same reparametrisation of the previous methods we obtain the following equation to learn the shifted mean: 
    
\[\mathbb{E}_{\epsilon, t, (x,y_0)}[C||\epsilon -\epsilon_\theta(\sqrt{\bar{\alpha}}y_0 + \sqrt{1-\bar{\alpha}_t}\epsilon, \tilde{x},\bar{\alpha_t}) + \frac{\Sigma}{C} \nabla_{y_t}||K*y_t - x||_m^m ||_2^2],\]

where \(C = \frac{\beta_t^2}{2\sigma_t\alpha_t(1-\bar{\alpha_t})}\), and \(\Sigma = \sigma_t^2 I\).

We also used the simplification of \(L_\text{simple}\) \cite{ho_denoising_2020} and trained by setting \(C = 1\). This allows us to have a weighted variational bound that emphasises different aspects of the image reconstruction. Moreover, removing this constant adds a smaller weight to terms corresponding to a smaller \(t\), which has been shown to be beneficial for higher-quality reconstructions. 

Furthermore, we re-weight the physics term by a constant sequence \(\nu_t\), such that the physics term remains relevant throughout the diffusion chain.
Thus, we train using the following simplified objective:

\[\mathbb{E}_{\epsilon, t, (x,y_0)}[||\epsilon -\epsilon_\theta(\sqrt{\bar{\alpha}}y_0 + \sqrt{1-\bar{\alpha}_t}\epsilon, \tilde{x},\bar{\alpha_t}) + \nu_t \nabla_{y_t}||K*y_t - x||_m^m ||_2^2].\]

This way, the diffusion model will predict the gradient of the physical problem alongside the \(\epsilon\) of \(L_\text{simple}\)The inference, in this case, is performed in the following way:

\[y_{t-1} = \frac{1}{\sqrt{\alpha_t}}\Bigg(y_t - \frac{\beta_t}{\sqrt{1-\bar{\alpha_t}}}\Bigg(\epsilon_\theta(x, y_t, \bar{\alpha_t}) +  \nu_t \nabla_{y_t}||K*y_t - x||_m^m\Bigg)\Bigg)+ \sqrt{\beta_t}\epsilon_t.\]

This is analogous to learning the unconditioned model and guide during the inference scaled by a specific constant depending on the schedule. This expression highlights the usefulness of our approach since it enables the incorporation of information only available during training to guide the DDPM inference. Additionally, we can add a regulariser in the same fashion as the model-based methods to have a DDPM model guided by an interpretable physical prior. 

\[y_{t-1} = \frac{1}{\sqrt{\alpha_t}}\Bigg(y_t - \frac{\beta_t}{\sqrt{1-\bar{\alpha_t}}}\epsilon(x, y_t, \bar{\alpha_t}) )\Bigg)+ \sqrt{\beta_t}\epsilon_t - \nu_t \frac{\beta_t}{\sqrt{1-\bar{\alpha_t}}} \nabla_{y_t}||K*y_t - x||_m^m- \lambda\nabla_{y_t}R(y_t).\]

\subsection{Metrics}
To assess the performance of all the models in this study we used multi-scale structural similarity index measure (MS-SSIM) \cite{wang2003multiscale}, normalised root mean square error \(\text{NRMSE}:=\sqrt{\text{MSE}}/y_{\text{max}}-y_{\text{min}}\), and peak single-to-noise ratio (PSNR). MS-SSIM metric is defined as:
\[\text{MS-SSIM}(x,y):= [l_M(x,y)]^{\alpha_M} \prod_{j=1}^M [c_j(x,y)]^{\beta_j} [s_j(x,y)]^{\gamma_j},\]
where \(l_j\), \(c_j\), and \(s_j\) are the measures of luminance, contrast, and structure corresponding to scale j. We used five scales and \(\alpha_j=\beta_j=\gamma_j\) for \(\sum_{j=1}^M \gamma_j = 1\) in accordance with the parameters reported in \cite{wang2003multiscale}. PSNR was defined as: 
\(\text{PSNR}:=20\log_{10}(\text{MAX}_I)-10\log_{10}(\text{MSE}),\)
where \(\text{MAX}_I\) is the maximum pixel value of the image and \(\text{MSE}\) is the mean square error.
\subsection{Training Details}
DDPM and PI-DDPM were trained in a cluster environment employing a single Nvidia A100 with 40GB of vRAM. The batch size was 16. Pre-training continued for 1 million iterations using an ImageNet-derived dataset and fine-tuned for 800 000 iterations on the mixed dataset. The U-Net model was trained in a cluster environment using a single Nvidia V100 GPU equipped with 32GB of vRAM. Pre-training continued for 80 000 iterations, and fine-tuning for 10 000. 
\section{Datasets}
\subsection{Simulated Datasets}
To train our models on the image reconstruction task for microscopy, we required a dataset large enough to prevent overfitting. For this we have constructed a simulated dataset using photography images  from ImageNet \cite{deng2009imagenet} containing 1.2 million training and 100 000 test images. To simulate micrographs, we have processed each image using a forward microscopy model, which we have termed the Image Acquisition Model (see Methods and Fig. 1c). We have then employed the processed ImageNet-derived dataset in accordance with the training and test holdouts defined in the original dataset and trained a U-Net \cite{ronneberger2015u}, DDPM \cite{ho_denoising_2020} models, as well as the PI-DDPM model proposed here on the task of reconstructing the high-resolution image from the simulated blurred micrograph.

\begin{wrapfigure}[30]{r}{0.5\textwidth}
  \begin{center}
    \includegraphics[width=0.48\textwidth]{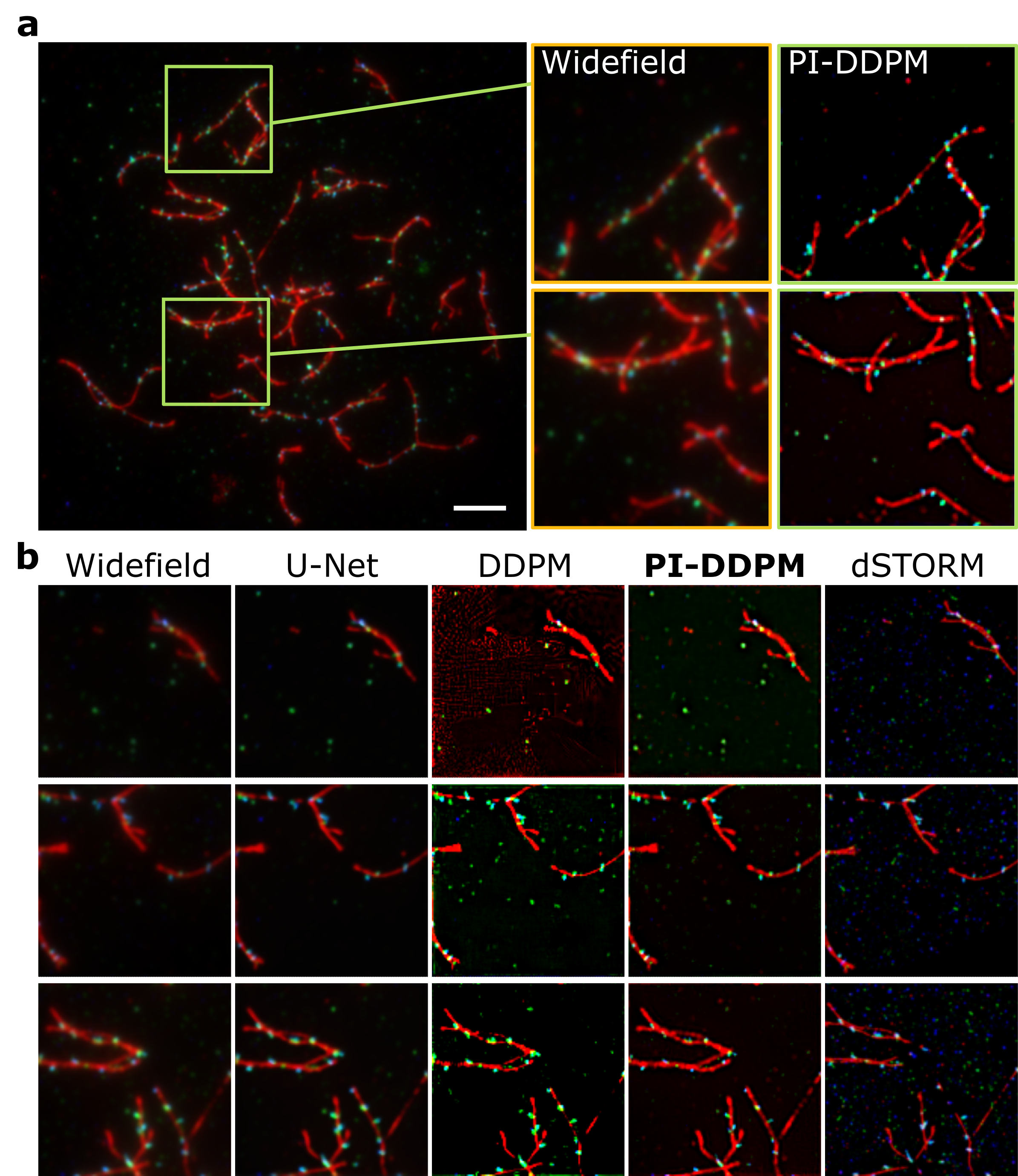}
  \end{center}
\caption{\begin{small}\textbf{Model performance on unseen superresolution dataset}. (a) image containing mid-zygotene nucleus immunostained for SYCP3 (red),  DMC1 (green) and RAD51 (blue) proteins from \cite{LiekeKoornneef2023}. (b) examples of input (Widefield) and reconstructed images using U-Net, Denoising Diffusion Probabilistic Models (DDPM) and physics-informed DDPM (PI-DDPM) in dSTORM images. The scale bar is 5 \(\mu\)m.\end{small}}
\end{wrapfigure}

Next, to further increase the relevance of our training dataset, we have combined the ImageNet-derived simulated microscopy dataset with a simulated dataset obtained from the publicly available BioSR dataset \cite{qiao2021evaluation}. BioSR dataset contains approximately 20 000 structured illumination microscopy (SIM) fluorescence images of subcellular structures like clathrin-coated pits (CCPs), endoplasmatic reticulum (ER), microtubules (MT) and F-actin imaged at varying levels of fluorescence. Specifically, from the BioSR containing pairs of low and high-resolution images we have taken all high-resolution images. We followed BioSR train-test split. During training, the high-resolution images constituted the ground truth, while images processed with our Image Acquisition Model (see Methods) we used as the input. This combined dataset was then used for fine-tuning our models. At test time we have used the BioSR low-resolution (widefield) images as input.
\subsection{Direct Stochastic Optical Reconstruction Microscopy Dataset}
To test how our method compares to the state-of-the-art single-molecule localisation microscopy (SMLM) we have employed a publicly available three-color Direct Stochastic Optical Reconstruction Microscopy (dSTORM) dataset \cite{LiekeKoornneef2023}. In this dataset authors provide a Widefield and SMLM reconstructed high-resolution image containing mid-zygotene nucleus immunostained for SYCP3 (red),  DMC1 (green) and RAD51 (blue) proteins. Images in this dataset were acquired by Zeiss Elyra PS1 microscope using a 100x 1.46NA oil immersion objective. Further imaging details are provided by the authors in the following publication \cite{koornneef2022multi}.

\subsection{Prospective Correlative Widefield-Confocal Microscopy Dataset}
Finally, to test how our model would perform on a prospectively acquired dataset, we have obtained a correlated widefield-confocal microscopy dataset. For this, A549 lung carcinoma cell line cells were seeded in 96-well imaging plates a night prior to imaging, then fixed with 4\% paraformaldehyde (Sigma) and stained for DNA with Hoechst 33342 fluorescent dye (Sigma). Cell culture was maintained similarly to the procedures described in \cite{yakimovich2015plaque2}. Next, stained cell nuclei were imaged using ImageXpress Confocal system (Molecular Devices) in either confocal or widefield mode employing Nikon 20X Plan Apo Lambda objective. To obtain 3D information images in both modes were acquired as Z-stacks with 0.3 \(\mu\)m and 0.7 \(\mu\)m for confocal and widefield modes respectively. Confocal z-stack was Nyquist sampled. The excitation wavelength was 405 nm and the emission 452 nm. Using these settings we have obtained 72 individual stacks for both modalities, with each stack covering 2048 by 2048 pixels or 699 by 699 \(\mu\)m.

\section{Experiments and Results}
\subsection{Training, Testing and Comparison of Physics-informed Denoising Diffusion Probabilistic Model using Simulated Microscopy Dataset}

To train the Physics-informed Denoising Diffusion Probabilistic Model (PI-DDPM) we proposed, we have employed a simulated dataset microscopy dataset (see Datasets). For this, we have processed ImageNet/BioSR-derived images using our image acquisition model (IAM, see Methods). These images used for training bore a strong resemblance to images obtained using LM. Changes that such processing inflicts on the images are demonstrated in Fig. 2a.
\begin{table}[ht]
  \caption{Performance on Widefield Microscopy using BioSR Test Set}
  \label{table}
  \centering
  \begin{tabular}{lllll}
    \toprule
    Metric / Model & Original & U-Net & DDPM & PI-DDPM (ours) \\
    \midrule
    PSNR & 18.649 & 19.706 & 23.703 & \textbf{23.974} \\
    MS-SSIM & 0.628 & 0.652 & 0.784 & \textbf{0.795} \\
    NRMSE & 0.147 & 0.126 & 0.070 & \textbf{0.069} \\
    \bottomrule
  \end{tabular}
\end{table}
To compare the performance of our PI-DDPM to other models, alongside PI-DDPM we have trained a U-Net\cite{ronneberger2015u}, DDPM\cite{ho_denoising_2020}. Additionally, for comparison, we added the model-based Richardson-Lucy (RL) \cite{richardson1972bayesian,lucy1974iterative,dey_richardsonlucy_2006} algorithm (Fig. 2b,c). The visual comparison of the model's performance suggested that DDPM and PI-DDPM showed less noise and processing artefacts in comparison to RL and U-Net. Furthermore, PI-DDPM preserved much more high-frequency details in both ImageNet and BioSR-derived images. To obtain quantitative performance measurements, we have computed the performance metrics including PSNR, MS-SSIM, and NRMSE (see Supplementary Materials). These metrics suggest that DDPM and PI-DDPM perform best among the algorithms we compared, while both models performed comparably.

Next, to test how the models perform on the reconstruction task from real rather than simulated microscopy we ran inference of U-Net, DDPM and PI-DDM on widefield images of the BioSR test set (Tab. 1). Both DDPM and PI-DDPM outperformed U-Net in all three metrics. Remarkably, on this real microscopy set PI-DDPM outperformed DDPM in all three metrics.

\subsection{Testing Physics-informed Denoising Diffusion Probabilistic Model using Direct Stochastic Optical Reconstruction Microscopy Dataset}

To test how PI-DDPM would perform on a previously unseen microscopy dataset we have employed a publicly available Direct Stochastic Optical Reconstruction Microscopy Dataset (dSTORM) \cite{LiekeKoornneef2023}. To circumvent pixel shift between widefield image and processed image we have chosen regions in the centre of the image where the shift was minimal. Remarkably, PI-DDPM produced results visibly consistent with the ground truth (Fig. 3a). In comparison to the U-Net, both DDPM and PI-DDPM produced visibly sharper reconstructions (Fig. 2b). Remarkably, PI-DDPM produces visibly less artefacts in low signal (red channel, upper row) compared to DDPM. Additionally, PI-DDPM preserved the continuity of the filament structures better (red channel, lowest row) compared to DDPM. Interestingly, DDPM seems to overemphasise the green channel (DMC1) possibly due to the high signal-to-noise. Noteworthy, in comparison to the dSTORM reconstruction all models failed to capture the low signal-to-noise punctae in RAD51 (blue channel). We have next quantified the performance of all the models for each channel and averaged the results (Tab. 2). Results show that both DDPM and PI-DDPM perform best in all three metrics. Remarkably, PI-DDPM outperformed DDPM in PSNR and NRMSE metrics. Furthermore, PI-DDPM performed comparably to DDPM in MS-SSIM.
\begin{table}[ht]
  \caption{Performance on dSTORM Test Set. Error stands for standard deviation.}
  \label{table}
  \centering
  \begin{tabular}{lllll}
    \toprule
    Metric / Model & Original & U-Net & DDPM & PI-DDPM (ours) \\
    \midrule
    PSNR & 16.487 & 16.712 & 15.541\(\pm\)0.232 & \textbf{16.778\(\pm\)0.807} \\
    MS-SSIM & 0.293 & 0.479 & \textbf{0.638\(\pm\)0.007} & \textbf{0.612\(\pm\)0.039} \\
    NRMSE & 0.150 & 0.146 & 0.167\(\pm\)0.005 & \textbf{0.145\(\pm\)0.015} \\
    \bottomrule
  \end{tabular}
\end{table}
\subsection{Testing Physics-informed Denoising Diffusion Probabilistic 
Model using Prospective Correlated Widefield-Confocal Microscopy Dataset}
\begin{wrapfigure}[33]{r}{0.5\textwidth}
  \begin{center}
    \includegraphics[width=0.48\textwidth]{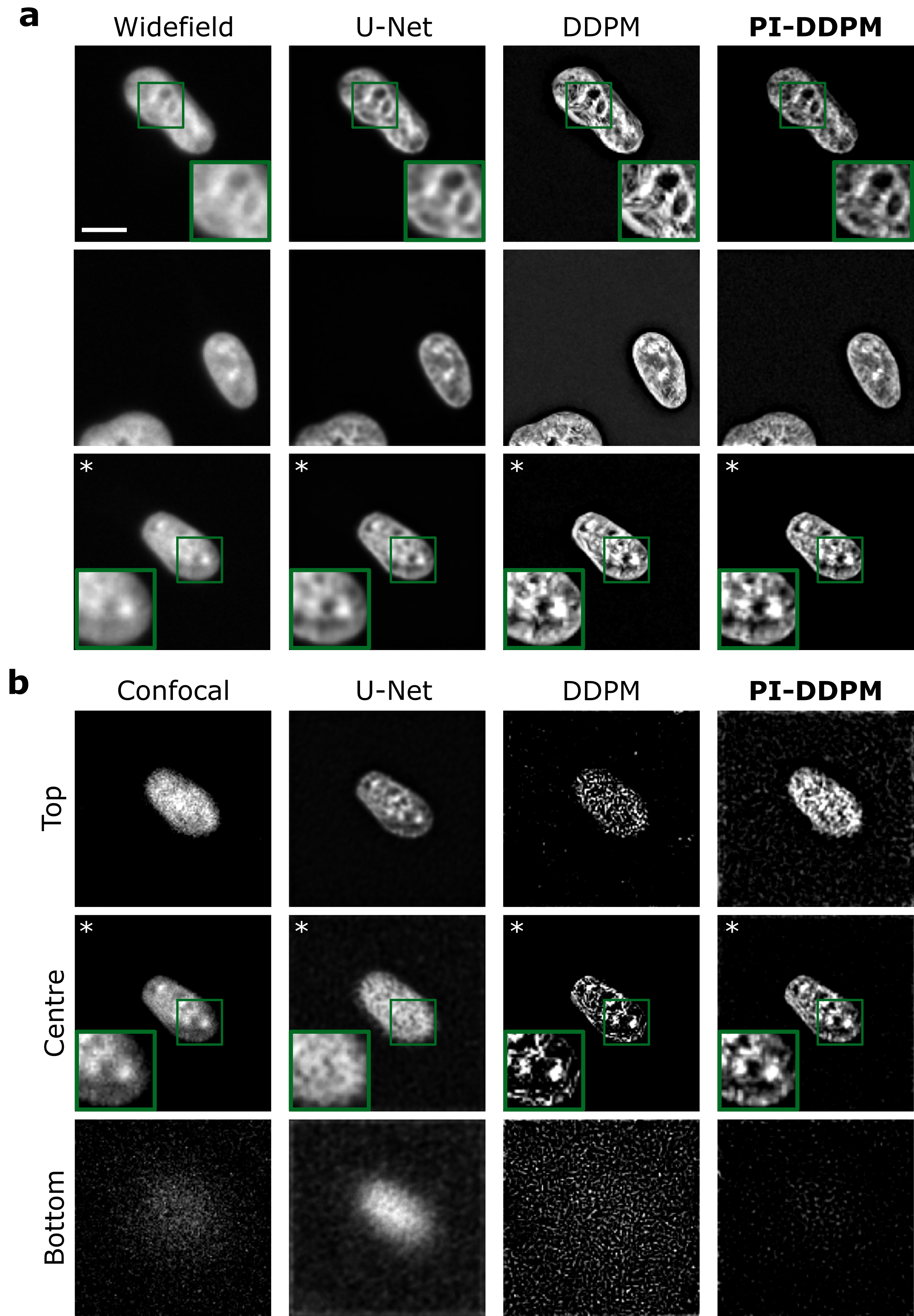}
  \end{center}
\caption{\begin{small}\textbf{Model performance on prospective correlative widefield-confocal microscopy}. (a) examples of widefield images of cell nuclei and their reconstructions using U-Net, Denoising Diffusion Probabilistic Models (DDPM) and physics-informed DDPM (PI-DDPM). (b) examples of confocal images and their reconstructions. The scale bar is 10 \(\mu\)m. Asterisk (*) marks correlated images of the same cell and focal plane.\end{small}}
\end{wrapfigure}
Finally, to test our image reconstruction model on a dataset acquired prospectively, we obtained a correlative widefield-confocal microscopy dataset of cell nuclei (Fig. 4). In this dataset, both confocal and widefield stacks of the same region were taken by automated microscopy. Since confocal microscopy (Fig. 4b) is known to have better resolution compared to widefield microscopy (Fig. 4a) it can serve as a guide on the correctness of image restoration in the absence of \textit{bona fide} ground truth. Consistent with our previous observations, we noted that DDPM and PI-DDPM have shown significantly lower blur in the reconstructions (Fig. 4a, top row).

However, compared to PI-DDPM, the conventional DDPM had notable artefacts in chromatin structures distorting the resulting image. Furthermore, comparing correlated fields of view (Fig. 4a,b, white asterisk) PI-DDPM reconstructions show more consistency with the confocal image than the conventional DDPM or U-Net irrespective of whether the input image comes from widefield or confocal microscopy. Furthermore, PI-DDPM showed more consistent output in the case of a low signal-to-noise ratio (Fig. 4b, bottom).

\section{Limitations}
The performance of the model suggested here is similar to or better than that of the benchmarks. However, the quality of the output can be further improved in several ways. Firstly, due to the lack of available microscopy data, our models are trained on a simulated dataset. While the simulation model employs \textit{bona fide} complete physics understanding microscope image acquisition, this model can be further improved by incorporating specific physical parameters such as lens aberrations. Secondly, the performance could be improved by training the model on a larger amount of real microscopy data. Finally, training of DDPM and PI-DDPM models is computationally expensive and requires high-performance computing infrastructure.
\section{Discussion}
Despite the immense progress in microscopy, the ability to visualise the microscopic world remains limited due to hardware imperfections and physics boundaries \cite{nechyporuk2022principles}. While recent advances in deep learning and generative models promise to assist in overcoming these barriers, these models come with their own set of limitations.

Since their introduction, denoising diffusion probabilistic models (DDPMs) have shown great promise for generative modelling \cite{ho_denoising_2020,nichol2021improved,lugmayr2022repaint}. However, in the case of demanding applications like biomedical image restoration and superresolution microscopy, the absence of artefacts and hallucinations, as well as the correctness of produced structures are of paramount importance. In this work, we show that the introduction of a physical prior to a DDPM model can improve stability and generate more realistic reconstruction results. Our approach extends a recently introduced paradigm of physics-informed neural networks \cite{raissi2017physics} and extends the applicability of these methods to microscopy. Furthermore, since our models learn the distribution from the data, the mean and variance of the obtained reconstructions can be estimated directly. This, in turn, may provide a convenient way to obtain confidence of our reconstructions, which could facilitate broader adoption by the biomedical community.

\section*{Acknowledgments}
We thank Michael Hecht and his group for critical reading of this work. This work was partially funded by the Center for Advanced Systems Understanding (CASUS) which is financed by Germany’s Federal Ministry of Education and Research (BMBF) and by the Saxon Ministry for Science, Culture, and Tourism (SMWK) with tax funds on the basis of the budget approved by the Saxon State Parliament. MK was supported by the Heisenberg award from the DFG (KU 3222/2-1), as well as funding from the Helmholtz Association.

\bibliographystyle{unsrtnat}
\bibliography{refs}


\newpage
\section*{Supplementary Methods Description}

\subsection{Denoising Diffusion Probabilistic Models}

Denoising Probabilistic Diffusion Models (DDPMs) are latent variable models that use a sequence of latent variables to model the data \cite{ho_denoising_2020}. These models use a Markov chain in which noise is gradually used to diffuse the data sample signal. This process is usually called the forward process. Formally, given $x_0 \sim q(x_0)$, the data distribution. We define a Markovian chain $q$ with states $x_{1:T}$, adding noise at each state according to a variance schedule $\beta_{1:T}$. The transitions of the Markov chain are defined as follows:

\[q(x_t|x_{t-1}):= \mathcal{N}(x_t; \sqrt{1-\beta_t}x_{t-1}, \beta_tI)\]

This definition allows computing $q(x_t|x_0)$, which can be expressed as the following normal distribution:

\[q(x_t|x_0)= \mathcal{N}(x_t; \sqrt{\Bar{\alpha_t}}x_0, (1-\Bar{\alpha_t})I)\]

\centerline{where $\alpha_t = 1 - \beta_t$ and $\Bar{\alpha_t} = \prod_{s=0}^t\alpha_s$.}

Then we can find the posterior $q(x_{t-1}|x_t,x_0)$ using the Bayes Theorem. The posterior is:

\[q(x_{t-1}|x_{t}, x_0)= \mathcal{N}(x_{t-1}; \Tilde{\mu_t}(x_t,x_0),\Tilde{\beta_t}I)\]

where the mean $\Tilde{\mu_t}$ is:
\[\Tilde{\mu_t}(x_t,x_0):=\frac{\sqrt{\Bar{\alpha}_{t-1}}\beta_t}{1-\Bar{\alpha_t}} x_0 + \frac{\sqrt{\alpha_t}(1-\Bar{\alpha}_{t-1})}{1-\Bar{\alpha}_{t}}x_t \]

and the variance is:
\[\Tilde{\beta_t} = \frac{1-\Bar{\alpha_{t-1}}}{1-\Bar{\alpha_{t}}}\beta_t\]

Then to sample from $q(x_0)$, we can sample from $q(x_T)$ which under sensible parameter selections $\beta_{1:T}$ and $T$, approaches a normal distribution $\mathcal{N}(x_T; 0,I)$ \cite{dhariwal_diffusion_2021} and then follow the reverse process $q(x_{t-1}|x_t)$. However, since we do not know the distribution $q(x_0)$, we train a neural network to approximate the reverse transition probability distribution. This function approximates a normal diagonal distribution \cite{sohl2015deep} in which, as the number of steps approaches infinity, the covariance matrix norm approaches $0$.
Then we only need to predict each timestep's mean and variance.
Formally, the learnable transition function is defined as follows:
\[p_\theta(x_{t-1}|x_{t}) = \mathcal{N}(x_{t-1};\mu_\theta(x_t,t), \Sigma_\theta(x_t,t))\]
Lastly, we can solve a lower-bound variational problem to optimise the model. However, it is possible to train the model by matching the mean of the posterior of the forward process and the mean of the reverse process:
\[\min_\theta \mathbb{E}_{t,x_t}||\mu_\theta(x_t,t) - \Tilde{\mu_t}(x_t,x_0)||\]

Using the reparametrisation

\begin{align*}
q(x_t|x_0)&= \mathcal{N}(x_t; \sqrt{\Bar{\alpha_t}}x_0, (1-\Bar{\alpha_t})I)\\
 	&= \sqrt{\Bar{\alpha_t}}x_0 + \epsilon\sqrt{1-\Bar{\alpha_t}}, \quad \epsilon\sim \mathcal{N}(\epsilon;0,I)
\end{align*}

\[\mu(x_t, x_0) = \frac{1}{\sqrt{\alpha_t}}\Bigg(x_t - \frac{\beta_t}{\sqrt{1-\Bar{\alpha}_t}}\epsilon \Bigg)\]
\[\mu_\theta(x_t, t) = \frac{1}{\sqrt{\alpha_t}}\Bigg(x_t - \frac{\beta_t}{\sqrt{1-\Bar{\alpha}_t}}\epsilon_\theta(x_t,t)\Bigg)\]

 The model is trained to predict $\epsilon$ at each timestep.
\[\min_\theta \mathbb{E}_{t,x_0,\epsilon}||\epsilon - \epsilon_\theta(x_t,t)||_2^2.\]

For the variance, it is suggested to use a constant such as $\beta_tI$ or $\Tilde{\beta_t}I$, which corresponds to the upper and lower bounds for the true variance of the reverse process \cite{ho_denoising_2020}. 

\subsection{Source Code}
The source code for this work is available at \url{https://github.com/casus/pi-ddpm}. To train the model place your data generated by the dataset\_generation script (if you are generating simulated data) or the STORM script if you are generating the respective STORM dataset.
In the train\_ddpm or train\_unet script change the paths of the loading data to the ones that you generated. Next, choose a training modality, either widefield or confocal. Finally, run the script.

To test the model generate your testing dataset using the dataset\_generation script.
Change the paths corresponding to your data. Next, change the paths to the weights files that you wish to use. Finally, run the test script.

Due to the size limitation of the submission for the prospective dataset containing correlative widefield-confocal fluorescence microscopy, we provide only a single stack as an example (data/teaser\_c\_w\_test.npz). The full dataset will be made available under CC-BY license upon acceptance of the manuscript.

\section{Supplementary Results}

\subsection{Simulated Microscopy Dataset}
\begin{table}[ht]
  \caption{[Supplementary] Performance on Simulated Microscopy using BioSR Test Set}
  \label{table}
  \centering
  \begin{tabular}{llllll}
    \toprule
    \cmidrule(r){1-2}
    Metric / Model & Original & Richardson-Lucy & U-Net & DDPM & PI-DDPM \\
    \midrule
    PSNR & 16.127 & 13.394 & 18.301 & \textbf{20.446} & \textbf{20.217} \\
    MS-SSIM & 0.745 & 0.684 & 0.812 & \textbf{0.859} & \textbf{0.859} \\
    NRMSE & 0.156 & 0.214 & 0.122 & \textbf{0.095} & \textbf{0.098} \\
    \bottomrule
  \end{tabular}
\end{table}

To train our models on the image reconstruction task for microscopy, we required a dataset large enough to prevent overfitting. For this we have constructed a simulated dataset using photography images  from ImageNet \cite{deng2009imagenet} containing 1.2 million training and 100 000 test images. To simulate micrographs, we have processed each image using a forward microscopy model, which we have termed the Image Acquisition Model (see Methods and Fig. 1c). We have then employed the processed ImageNet-derived dataset in accordance with the training and test holdouts defined in the original dataset and trained a U-Net \cite{ronneberger2015u}, DDPM \cite{ho_denoising_2020} models, as well as the PI-DDPM model proposed here on the task of reconstructing the high-resolution image from the simulated blurred micrograph. Supplementary Table 1 contain test-time performance using multi-scale structural similarity index measure (MS-SSIM) \cite{wang2003multiscale}, normalised root mean square (NRMSE), and peak single-to-noise ratio (PSNR) metrics.

\newpage
\begin{figure}
  \begin{center}
    \includegraphics[width=1\textwidth]{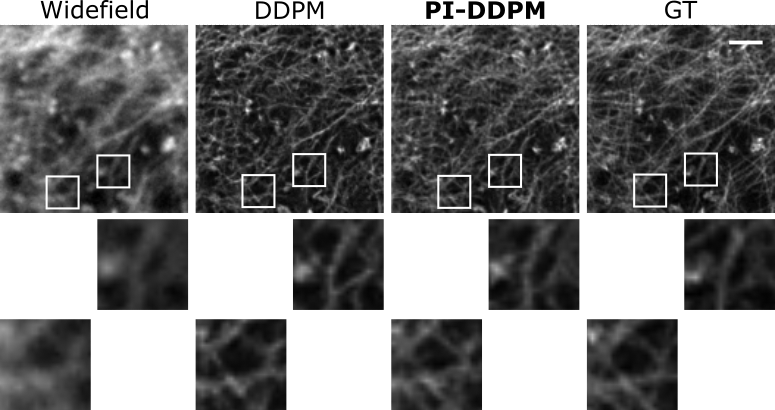}
    \caption{\begin{small}\textbf{[Supplementary] Comparison of model performance on BioSR-derrived dataset}. Denoising Diffusion Probabilistic Models (DDPM) and physics-informed DDPM (PI-DDPM) in BioSR images. GT stands for ground truth. MT stands for microtubules. Scale bar in micrographs is 1.5 \(\mu\)m.\end{small}}
  \end{center}

\end{figure}

Similarly, to the results shown in the main manuscript, these test results suggest that the performance of DDPM and PI-DDPM models surpassed all the other models. However, upon closer inspection of the reconstructed structures, DDPM shows a significant amount of hallucinated structures (Sup. Fig. 1). Specifically in the top row of zoomed-in insets closing of the filament structure may be observed in DDPM, which is absent in both PI-DDPM and ground truth (GT). The examples shown in the lower row of insets demonstrate an opposite hallucination. A structure that is meant to be continuous according to the ground truth and PI-DDPM, appears broken in DDPM reconstruction.


\end{document}